\documentclass[superscriptaddress,notitlepage,nofootinbib,a4paper,11pt]{revtex4-1}
\usepackage[UKenglish]{babel}
\usepackage{dsfont}
\usepackage{amssymb}
\usepackage{mathrsfs}
\usepackage{amsmath}
\usepackage{amsthm}
\usepackage{amsfonts}
\usepackage{mathtools}
\usepackage{footnote}
\usepackage{graphicx}
\usepackage{verbatim}
\usepackage{bbm}
\usepackage{tikz}
\usepackage{verbatim}
\usepackage{pgfplots}
\usepackage{gensymb}
\usepackage[left=2.5cm,right=2.5cm]{geometry}
\usepackage[hyperfootnotes=false]{hyperref}
\usepackage{natbib}
\usepackage{fancyhdr} 
\usepackage{accents}

\fancypagestyle{newstyle}{
\fancyhf{} 
\fancyfoot[R]{\thepage\centering} 

}
\numberwithin{equation}{section}

\newtheorem{thm}{Theorem}[section]
\newtheorem{lem}[thm]{Lemma}

\newtheorem{cor}[thm]{Corollary}
\newtheorem{prop}[thm]{Proposition}

\newtheorem*{lem*}{Lemma}

\newcommand\ve{\varepsilon}
\newcommand\vf{\varphi}
\newcommand\nn{\nonumber}

\newcommand{\IIp}{[-\ve^{-1},\ve^{-1}]}

\newcommand{\vei}{\ve^{-1}}
\newcommand{\veim}{\ve^{-{1\over 2}}}
\newcommand{\de}{\mathrm{d}}

\newcommand{\sqve}{\sqrt{\ve}}

\newcommand{\mb}{m_{\beta}}
\newcommand{\ms}{m^{*}\left(\beta\right)}
\newcommand{\chib}{\chi_{\beta}}

\newcommand{\hs}{\hspace*{0.4mm}}

\begin{document}

\title{\normalfont{\Large Stationary currents in long-range interacting magnetic systems}}
\author{Roberto Boccagna}
\affiliation{{Universit{\`a} dell'Aquila, Via Vetoio, Loc. Coppito, 67010 L'Aquila, Italia.}}

\begin{abstract}
\noindent
We construct a solution for the $1d$ integro-differential stationary equation derived from a finite-volume version of the mesoscopic model proposed in \cite{GIAC}. This is the continuous limit of an Ising spin chain interacting at long range through Kac potentials. The microscopic system is in contact with reservoirs of fixed magnetization and infinite volume, so that their density is not affected by any exchange with the bulk in the original Kawasaki dynamics. At the mesoscopic level, this condition is mimicked by the adoption Dirichlet boundary conditions. We derive the stationary equation of the model starting from the Lebowitz-Penrose free energy functional defined on the interval $[-\vei,\vei]$, $\ve>0$. For $\ve$ small, we prove that below the critical temperature there exists a solution that carries positive current provided boundary values are opposite in sign and lie in the \textit{metastable} region. Such profile is no longer monotone, connecting the two phases through an antisymmetric interface localized around the origin. This represents an analytic proof of the existence of diffusion along the concentration gradient in one-component systems undergoing a phase transition, a phenomenon generally known as \textit{uphill diffusion}. However uniqueness is lacking, and we have a clue that the stationary solution obtained is not unique, as suggested by numerical simulations.
\end{abstract}

\maketitle

\noindent
\textbf{Keywords}: uphill diffusion, Kac potentials, Fick's law, phase transitions

\section{Introduction}

The aim of this paper is to study Fick's law of transport in one-component systems undergoing a second order phase transition. In this context, it represents a step forward towards the establishment of a well posed theory for diffusion along the gradient (\textit{uphill diffusion}). Fick's law relates the flux $J$ of a given substance to the gradient of its concentration $\rho$, which we suppose to be a differentiable function of the position in $\left[0,L\right]$:
\begin{equation}\label{fick}
J=-D\,\frac{\de}{\de x}\rho,
\end{equation}
at fixed boundary conditions $\rho\left(0\right)=\rho_-$, $\rho\left(L\right)=\rho_+$, with $\rho_-<\rho_+$ WLOG. Here, $D>0$ is the diffusion coefficient. According to \eqref{fick}, the flux is always in the direction of decreasing gradient, i.e. from the region at higher concentration to the region at lower concentration. Thus, the solution of \eqref{fick} connects monotonically $\rho_-$ to $\rho_+$, as represented in a sketchy way in Figure \ref{diff1}. Indeed, \eqref{fick} should be modified when considering systems that consist of many components, since diffusion may be also affected by possible microscopic, chemical interactions among different substances. Evidences of surprising behaviors have already been reported by Nernst \cite{NERNST}, Onsager \cite{ONS} and especially Darken \cite{DARKEN1,DARKEN2,LSD}, who performed an acknowledged experiment in the late 40's. His setup consisted of pairs of doped steels (Fe-Si with a different wt. $\%$ of silicon, Fe-Si and Fe-Mn or Fe-Si and Fe-Mo) containing a small difference in the carbon concentration at the edges. The steels were welded together and eventually held in a furnace in order to let diffusion occur. In fact, it was observed that carbon diffused following the gradient in the mixtures with slightly differences in carbon concentration. This is shown in Figure \ref{diffu}, which refers to the Fe-Si-Mn compound after two weeks the experiment started. 

%%%%%%%%%%%%%%%%%%%%%%%%%%%%%%%%%%%%%%%%%%%%%%%%%%%%%%

\begin{figure}[htbp!]
\centering
\begin{tikzpicture}[trim axis left, trim axis right]
\begin{axis}[
width=10cm,
height=4.9cm,
scale only axis,
axis lines=middle,
inner axis line style={=>},
ylabel=$$,
xlabel=$x$,
xmin=-1,
xmax=4,
ymin=0,
ymax=1,
domain=-0.2:4.1,
xtick={10},
ytick={5},
extra x ticks={0,3},
extra x tick labels={$0$,$L$},  
]
\draw[thick,color=black] (axis cs:-1,0.2) -- (axis cs:0,0.2);
\draw[color=black] (axis cs:3,0) -- (axis cs:3,1);
\draw[dashed,color=black] (axis cs:0,0.8) -- (axis cs:3,0.8);
\draw[thick,color=black] (axis cs:3,0.8) -- (axis cs:4,0.8);
\draw[thick,color=black] (axis cs:0,0.2) -- (axis cs:3,0.8);
\node[] at (80,80)   (a) {$\rho_L$};
\node[] at (80,13)   (a) {$\rho_0$};
\node[] at (245,60)   (a) {$J<0$};
\end{axis}
\end{tikzpicture}
\caption{Sketchy representation of the solution of \eqref{fick}.}
\end{figure}
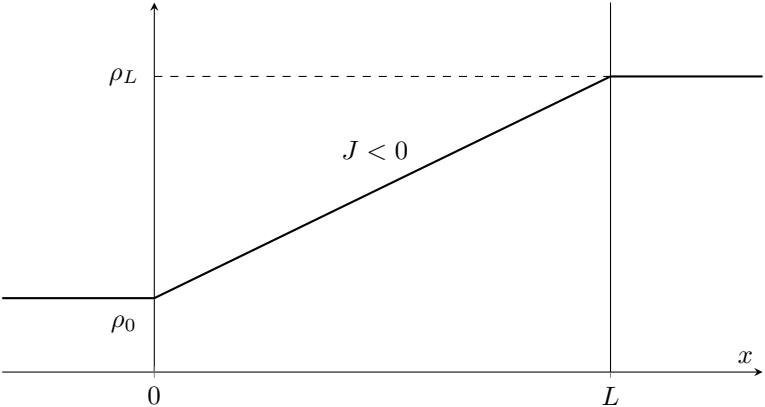\label{diff1}

%%%%%%%%%%%%%%%%%%%%%%%%%%%%%%%%%%%%%%%%%%%%%%%%%%%%%%

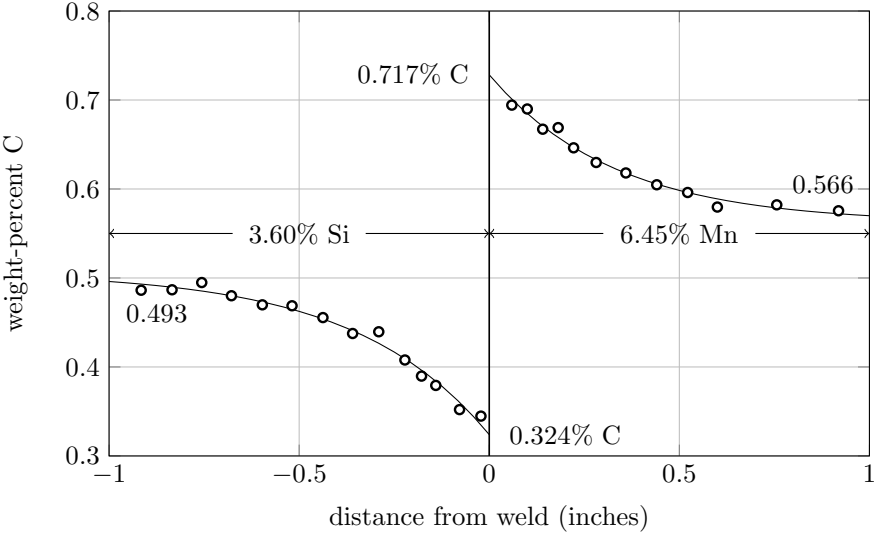
\begin{figure}[htbp!]
\centering
\begin{tikzpicture}[trim axis left, trim axis right]
\begin{axis}[
grid=both,
width=10cm,
height=5.9cm,
scale only axis,
ylabel=$\text{weight-percent C}$,
xlabel=$\text{distance from weld (inches)}$,
ymin=0.30,
ymax=0.80,
xmin=-1,
xmax=1,
xtick={-1.0,-0.5,0,0.5,1.0},
ytick={0.3,0.4,0.5,0.6,0.7,0.8},
]
\addplot[only marks,mark=*,mark size=0.6mm,mark options={color=black,line width=0.9pt,fill=white},color=black]
table{SiMn.dat};
\addplot[domain=0.001:1]{0.561987+0.166301*exp(-3.03827*x)};
\addplot[domain=-1:0.001]{0.506756-0.182751*exp(2.83615*x)};
\draw[<-,thin] (axis cs:-1,0.55) -- (axis cs:-0.675,0.55);
\draw[->,thin] (axis cs:-0.325,0.55) -- (axis cs:0,0.55);
\draw[<-,thin] (axis cs:0,0.55) -- (axis cs:0.3,0.55);
\draw[->,thin] (axis cs:0.7,0.55) -- (axis cs:1,0.55);
\draw[black,semithick] (axis cs:0,0.3) -- (axis cs:0,0.8);
\node[] at (axis cs: -0.5,0.55) {3.60$\%$ Si};
\node[] at (axis cs: 0.5,0.55) {6.45$\%$ Mn};
\node[] at (axis cs: -0.2,0.73) {0.717$\%$ C};
\node[] at (axis cs: 0.2,0.324) {0.324$\%$ C};
\node[] at (axis cs: -0.875,0.46) {0.493};
\node[] at (axis cs: 0.8775,0.605) {0.566};
\end{axis}
\end{tikzpicture}
\caption{Carbon concentration in the Fe-Si-Mn compound after 10 days at 1050 $\degree$C (figure extrapolated from \cite{LSD}).}
\end{figure}\label{diffu}

%%%%%%%%%%%%%%%%%%%%%%%%%%%%%%%%%%%%%%%%%%%%%%%%%%%%%%

This counterintuitive behavior has a microscopic origin: in fact, silicon decreases the chemical affinity of carbon, while manganese increases it. This results in a driving force that acts in the opposite direction with respect to the concentration gradient and it might beat the gradient, if the difference in the carbon concentration at the edges is small. Such mechanism, which actually sustains uphill diffusion, works until dopants penetrate the weld; then, ``standard'' diffusion regime is restored. In formulae, this can be modeled by replacing \eqref{fick} with a vectorial relation:
\begin{equation}\label{fick2}
J_i=-c_i \sum_j L_{ij}\nabla \mu_j, \qquad i=1,\ldots,n
\end{equation}
where $n>2$ is the number of components of the system, $i$ refers to a given component, $c_i$ is the molar concentration of $i$, $L_{ij}$ represents the Fick diffusivity of $i$ given the presence of $j$, while the $\mu_i$'s are chemical potentials. \eqref{fick2} describes a system of $n-1$ linearly independent equations, because of the Gibbs-Duhem relation for chemical potentials \cite{KR1,KR2}. 
\newline
Very surprisingly, numerical simulations suggest that uphill diffusion may also occur also in one-component systems undergoing a phase separation. Colangeli et al. considered in \cite{CDMP} a $1d$ stochastic automaton describing a dissipative system of particles interacting at large distances. After a transient, a stationary state with non zero current emerges and, moreover, a region in which diffusion follows the concentration gradient can be spotted tuning the characteristic parameters of the system. Similar results have been obtained by running a Kawasaki dynamics for an Ising spin chain with Kac potentials below the critical temperature, in which particles located at the edges may flip according to assigned rates, in order to mimic interactions with reservoirs of infinite volume and opposite magnetization \cite{RB2}. When the magnetizations of the reservoirs are suitably chosen, the flux follows the ``magnetization gradient''. The resulting steady profile, called \textit{bump}, is no longer monotone and connects the two boundary values through an interface that is localized in the nearby of one of the edges, randomly selected by dynamics. Colangeli et al. \cite{CGGV} obtained analogue numerical results for the $2d$ nearest neighbors Ising model. 
\newline
The microscopic mechanism underlying uphill diffusion in one-component systems has not a chemical origin. We speculate that the ``force'' that counteracts the gradient is provided in this case by the separation of phases; however, we believe that such state is in fact metastable, meaning that bumps are local minima for the corresponding Gibbs free energy, but not global ones. Hence, after a transient, the flux should reverse to be directed from the state at higher magnetization to the state at lower magnetization. Nevertheless, such inversion does not take place in the time considered for simulations. 
\newline
\newline
\indent
Here we prove analytically the occurrence of uphill diffusion considering the model that is the continuous limit of the Ising chain with Kac potentials and Dirichlet boundary conditions. Our starting point is the Lebowitz-Penrose free energy functional, that is a non-local version of the scalar Ginzburg-Landau functional and that we postulate to describe the Physics of the system at the \textit{mesoscopic} level. This represents the intermediate scale between the microscopic, discrete chain and the macroscopic model, which is obtained letting the size of the system diverge. We know that in this case the phase diagram (i.e. the free energy density vs. magnetization diagram) has a global minimum for $\beta<1$, $\beta$ the inverse temperature in our units, while for $\beta>1$ the graph is flat in $\left[-\mb,\mb\right]$, $\mb$ the positive solution of the mean field equation $m=\tanh\left(\beta m\right)$. This indicates the occurrence of a phase transition. Any value in $\left(-\mb,\mb\right)$, the so called \textit{spinodal region}, is then forbidden, so that any stationary profile containing values smaller of $-\mb$ and  larger than $\mb$ must be discontinuous. At the mesoscopic level, the spinodal region is actually available and the discontinuity is replaced by a smooth interface, as proved by De Masi et al. for the free-boundary Stefan problem \cite{EP1,EP2,EP3,EP4}. However, the discontinuity is recovered when the hydrodynamic limit is performed.
\newline
We then fix $\beta>1$ ($\beta=1$ the critical inverse temperature in the mean-field model) and look for stationary solutions of $\dot{m}=-\frac{\partial }{\partial x} I$ in the space of bounded antisymmetric functions, $I$ being the local current which is supposed to be proportional to the functional derivative of the free energy. Thus, we reduce to the problem $\frac{\partial }{\partial x} I=0$ in the finite interval $[-\vei,\vei]$, $\ve>0$ fixed, which turns out to be an integro-differential equation \cite{GIAC}. The corresponding Dirichlet problem has been already studied in \cite{RB2} and \cite{RB}, although in the presence of an external, antisymmetric magnetic field. In that case it has been proved that, whatever the intensity of this field, the provided ``external force'' cannot reverse the flux when the positive boundary condition is in the interval $\left(\ms,\mb\right)$, $\ms=\sqrt{1-1/\beta}$ the positive saddle point of the mean field free energy (Figure \ref{dyy}). 

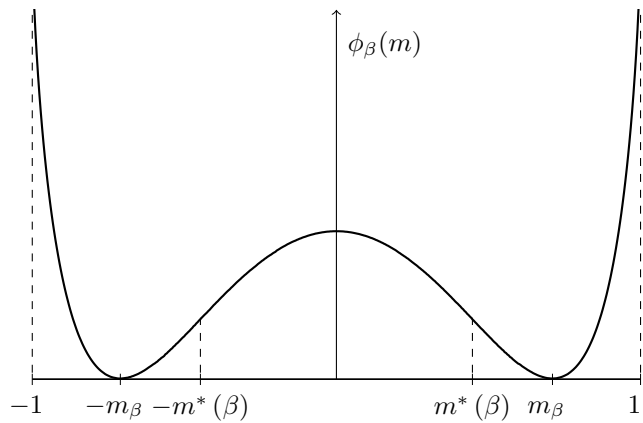
\begin{figure}[htbp!]
\centering
\begin{tikzpicture}[trim axis left, trim axis right]
\begin{axis}[
white,
every x tick/.style={black},
width=10cm,
height=4.9cm,
scale only axis,
axis lines=middle,
inner axis line style={=>},
xlabel=$m$,
xmin=-1.25,
xmax=1.25,
ymin=0.638,
ymax=0.71,
domain=-1.1:1.1,
xtick={5},
ytick={-1},
extra x ticks={-1,-0.710412,-0.4472,0.4472,0.710412,1},
extra x tick labels=
{$\textcolor{black}{-1}^{\textcolor{white}{*}}$,
$\textcolor{black}{-m_{\beta}}^{\textcolor{white}{*}}$,
$\textcolor{black}{-m^*\left(\beta\right)}$,
$\textcolor{black}{m^*\left(\beta\right)}$,
$\textcolor{black}{m_{\beta}}^{\textcolor{white}{*}}$,
$\textcolor{black}{1}^{\textcolor{white}{*}}$},  
]
\draw[dashed,color=black] (axis cs:0.4472,0.63) -- (axis cs:0.4472,0.65);
\draw[dashed,color=black] (axis cs:-0.4472,0.63) -- (axis cs:-0.4472,0.65);
\draw[dashed,color=black] (axis cs:1,0.63) -- (axis cs:1,0.8);
\draw[dashed,color=black] (axis cs:-1,0.63) -- (axis cs:-1,0.8);
\draw[thick,color=black] (axis cs:-0.4472,0.638) -- (axis cs:0.4472,0.638);
\draw[thick,color=black] (axis cs:-0.71,0.638) -- (axis cs:-0.4472,0.638);
\draw[thick,color=black] (axis cs:0.71,0.638) -- (axis cs:0.4472,0.638);
\draw[thick,color=black] (axis cs:0.71,0.638) -- (axis cs:1,0.638);
\draw[thick,color=black] (axis cs:-0.71,0.638) -- (axis cs:-1,0.638);
\draw[black,->] (axis cs:0,0.638) -- (axis cs:0,0.71);
\node[] at (141,652)   {$\textcolor{black}{\phi_{\beta}(m)}$};
\addplot[thick,color=black,samples=500]
{-(x^2/2)+0.8*((1-x)/2)*ln((1-x)/2)+0.8*((1+x)/2)*ln((1+x)/2)+1.2214};
\end{axis}
\end{tikzpicture}
\caption{The mean field free energy at $\beta>1$.}
\end{figure}\label{dyy}

We solve the stationary equation and prove that our resulting profile actually carries positive current. For $x\ge 0$ the solution firstly increases, jumping ``instantly'' from zero to $m_{\beta}$, then decreases to the metastable value at the right boundary; however, the region in which the current flows in the ``wrong'' direction reduces to a set of zero Lebesgue measure in the hydrodinamic limit, so that Fick's Law actually holds almost everywhere (w.r.t. the Lebesgue measure). The weak spot of the analysis is that our solution is supposedly unstable; in fact, numerical simulations suggest that bumps should be stable points for the gradient dynamics.
\newline
\indent
The stationary Stefan problem in bounded domains has been already considered by De Masi, Presutti and Tsagkarogiannis in \cite{DMPT}, despite Neumann conditions have been adopted there. Apart from technicalities in the proof, the two  approaches are quite different, since in the Neumann setting the magnetization profile naturally selects the boundary values imposed by the choice of current. However, as clarified in that work, there are solutions of the mesoscopic Neumann version of problem that converge to any solution of the Dirichlet problem as $\ve\to 0$.

\section{Background}

\subsection{The Lebowitz-Penrose functional}

\noindent
Indicate for notational convenience $\Lambda\coloneqq\IIp$ and $\Lambda^c\coloneqq\mathbb{R}\setminus\IIp$. Consider $m_{\Lambda}\in L^{\infty}\left(\Lambda,\left[-1,1\right]\right)$, $m_{\Lambda^c}\in L^{\infty}\left(\Lambda^c,\left[-1,1\right]\right)$, $m_{\Lambda}$ being the magnetization density of the bulk and $m_{\Lambda^c}$ the magnetization of the reservoirs. Our starting point is the mesoscopic Lebowitz-Penrose free energy functional at zero external magnetic field, that is 
\begin{eqnarray}
&&\mathcal{F}_{\beta}\left[m_{\Lambda}\mid m_{\Lambda^c}\right] = \mathcal{F}_{\beta}\left[m_{\Lambda}\right] +{1\over 2} \int_{\Lambda}\int_{\Lambda^c}\tilde{J}\left(x,y\right)\left[m_{\Lambda}\left(x\right)-m_{\Lambda^c}\left(y\right)\right]^2\de x\,\de y, \label{freeen}\\
&&\mathcal{F}_{\beta}\left[m_{\Lambda}\right] = -{1\over\beta}\int_{\Lambda}\phi_{\beta}\left(m_{\Lambda}\left(x\right)\right)\de x + {1\over 4} \int_{\Lambda}\int_{\Lambda}\tilde{J}\left(x,y\right)\left[m_{\Lambda}\left(x\right)-m_{\Lambda}\left(y\right)\right]^2\de x\,\de y \label{freeen2}
\end{eqnarray}
where 
\begin{equation}
\phi_{\beta}\left(m\right)=-\frac{1}{2}m^2-\frac{1}{\beta}S\left(m\right),
\end{equation}
and $S\left(m\right)$ is the standard binary entropy for an Ising spin system:
\begin{equation}\label{entr}
S\left(m\right)=-{1+m\over 2}\log\left({1+m\over 2}\right)-{1-m\over 2}\log\left({1-m\over 2}\right).
\end{equation}
$\tilde{J}$ is a probability kernel that actually depends on the distance between two points. The assumptions made on $\tilde{J}$ are precisely listed below:
\begin{itemize}
\item[-] $\tilde{J}$ is translational invariant, that is $\tilde{J}\left(x,y\right)=\tilde{J}\left(0,\left|x-y\right|\right)$ for any $x,y\in\mathbb{R}$;
\item[-] $\tilde{J}$ is twice differentiable, compactly supported: $\tilde{J}\in C_K^{2}\left(\left[-1,1\right],\left[0,1\right]\right)$;
\item[-] $\tilde{J}$ is normalized, $\int_{\mathbb{R}} \tilde{J}\left(\,\cdot\,,y\right)\de y=1$;
\item[-] $\tilde{J}\left(0,x\right)$ is strictly decreasing in $\left[0,1\right]$.
\end{itemize}
We treat expressions \eqref{freeen}, \eqref{freeen2} as primitive quantities, by postulating them to define our model at a mesoscopic level. Indeed, it can be shown that this is precisely what one obtains taking the continuous limit of the underlying microscopic Ising chain with Kac potentials after a feasible scaling. We indicate reference \cite{GIAC} for details on this procedure, that is known as the \textit{Lebowitz-Penrose limit}. 
We drop hereafter the suffix $\Lambda$, also indicating $\mathcal{F}_{\beta}\left[m\mid \mu\right]\equiv\mathcal{F}_{\beta}\left[m_{\Lambda}\mid m_{\Lambda^c}\right]$ for the sake of simplicity.
\newline
The axiomatic theory provides that the magnetization evolves in time according to a gradient dynamics
\begin{equation}\label{motion}
\dot{m}\left(x,t\right)=-\frac{\partial I}{\partial x}\left(x,t\right), \qquad t\ge 0
\end{equation}
where $I$ represents the local current
\begin{equation}\label{stat}
I\left(x,t\right)\coloneqq-\chib\left(m\left(x,t\right)\right)\frac{\partial}{\partial x}\frac{\delta\mathcal{F}_{\beta,\mu}\left[m\mid \mu\right]
}{\delta m\left(x,t\right)},
\end{equation}
in which $\chib\left(m\right)=\beta\left(1-m^2\right)$ is the mobility coefficient for an Ising spin system. Hence, the stationary problem $\dot{m}=0$ reads
\begin{equation}\label{current}
I=-\chib\left(m\left(x\right)\right)\frac{\partial}{\partial x}\frac{\delta\mathcal{F}_{\beta}\left[m\mid \mu\right]
}{\delta m\left(x\right)},
\end{equation}
that is an integro-differential equation in the unknown function $m$ at constant $I$ and given boundary conditions. We call $I=j\ve$, $j\in\mathbb{R}$ a constant, as we expect the current to be of order $\ve$. 

\subsection{Notation}

\noindent
It is worth redefining the convolution kernel in the distributional sense as follows
\begin{equation}
J\left(x,y\right)\stackrel{{\mathcal{D}'}}{=}\tilde{J}\left(x,y\right)\mathbf{1}_{\left|y\right|<\vei}+b_{\ve}\left(\left|x\right|\right)\left[\delta(\vei-y)+\delta(\vei+y)\right]\mathbf{1}_{\left|y\right|\ge\vei},
\end{equation}
where
\begin{equation}
b_{\ve}\left(x\right)\coloneqq\int_{\vei}^{\vei+1}\tilde{J}\left(x,y\right)\de y.
\end{equation}
This way we act on functions defined in the interior of the bulk. For any bounded function in $[-\vei,\vei]$, we indicate $\left\|\,\cdot\,\right\|_{\ve} \equiv \left\|\,\cdot\,\right\|_{L^{\infty}\left([-\vei,\vei]\right)}$ as the sup norm in that interval. For any $m\in L^{\infty}\left([-\vei,\vei]\right)$, we define
\begin{equation}
\left(J\ast m\right)\left(x\right)\coloneqq \int J\left(x,y\right)m\left(y\right) \de y.
\end{equation}
If not specified, integrals are intended to be performed on $[-\vei,\vei]$. For any $m,h\in L^{\infty}\left([-\vei,\vei]\right)$ we call
\begin{equation}
p_{m,h}\left(x\right) \coloneqq \beta \cosh^{-2}\Big\{\beta\Big[\left(J\ast m\right)\left(x\right)+h\left(x\right)
\Big]\Big\}
\end{equation}
and
\begin{equation}
p'_{m,h}\left(x\right)\coloneqq p_{m,h}\left(x\right) \tanh\Big\{\beta\Big[\left(J\ast m\right)\left(x\right)+h\left(x\right)
\Big]\Big\}.
\end{equation}
Let $\mathsf{A}_{m,h}$ the linear operator acting on a bounded function $f$ as follows
\begin{equation}
\mathsf{A}_{m,h}f\left(x\right)= \int p_{m,h}\left(x\right) J\left(x,y\right)f\left(y\right) \de y.
\end{equation}
$\mathsf{A}_{m,h}$ is the linear operator with kernel $\mathsf{A}_{m,h}\left(x,y\right)=p_{m,h}\left(x\right)J\left(x,y\right)$. The action of the $n$-th power of $\mathsf{A}_{m,h}$ on $f$ is explicitly given by
\begin{equation}
\mathsf{A}_{m,h}^n f\left(x_0\right)=\int \prod_{i=1}^n p_{m,h}\left(x_{i-1}\right)J\left(x_{i-1},x_i\right)f\left(x_n\right)\de x_1\ldots\de x_n.
\end{equation}
 
\subsection{Instantons}

\noindent
We briefly recall a fundamental result obtained by De Masi et al. \cite{EP1,EP2,EP3,EP4}. This regards the free boundary version of the problem. Let 
\begin{equation}
\mathcal{F}_{\beta}\left[m\right] = -{1\over\beta}\int_{\mathbb{R}}\phi_{\beta}\left(m\left(x\right)\right)\de x + {1\over 4} \int_{\mathbb{R}} \int_{\mathbb{R}} J\left(x,y\right)\left[m\left(x\right)-m\left(y\right)\right]^2\de x\,\de y
\end{equation}
be the free energy functional on $\mathbb{R}$ defined for functions that belong to the Banach space
\begin{equation}
\mathcal{N}\coloneqq\Big\{m\in L^{\infty}\left(\mathbb{R},\left[-1,1\right]\right) \mid \liminf_{x\to -\infty}m\left(x\right)<0, \;  \limsup_{x\to +\infty}m\left(x\right)>0 \Big\}.
\end{equation}
In this case the flux is null, so that the stationary problem reduces to
\begin{equation}\label{evo}
\dot{m}\left(x,t\right)=-\frac{\delta\mathcal{F}_{\beta}\left[m\right]
}{\delta m\left(x,t\right)}, \qquad t\ge 0.
\end{equation}
Let the \textit{instanton} $\bar{m}$ the set of minimizers of $\mathcal{F}_{\beta}\left[m\right]$ in $\mathcal{N}$ solution of
\begin{equation}\label{inst}
\bar{m}\left(x\right)=\tanh\Big\{\beta\Big[\left(J\ast \bar{m}\right)\left(x\right)\Big]\Big\}, \qquad x\in\mathbb{R}
\end{equation}
that satisfies $\lim_{x\to\pm\infty}m\left(x\right)=\pm\mb$.
\newpage
\begin{thm}
For any $\beta>1$ the following holds:
\begin{itemize}
\item[-] the variational problem $\delta\mathcal{F}_{\beta}\left[m\right]=0$ has a minimizer which is unique up to translations;
\item[-] the mean field equation \eqref{inst} has a solution which is unique in $\mathcal{N}$, up to translations;
\item[-] for any $m\in\mathcal{N}$, there is $\xi\in\mathbb{R}$ such that $\lim_{t\to\infty} \left\|m\left(\,\cdot\,,t\right) -m_{\xi}\right\|_{\infty}=0$, where $\bar{m}_{\xi}\left(x\right)\coloneqq \bar{m}\left(x-\xi\right)$.
\end{itemize}
$\bar{m}\in C^{\infty}\left(\mathbb{R},\left[-1,1\right]\right)$ is a strictly increasing, antisymmetric function which converges exponentially fast to $\pm\mb$ as $x\to\pm\infty$.
\end{thm}
\noindent
Hereafter, we call
\begin{equation}
\bar{p}\left(x\right)\coloneqq \beta\left[1-\bar{m}^2\left(x\right)\right], \qquad {\mathsf{A}_{\bar{m}}}\left(x,y\right)\coloneqq \bar{p}\left(x\right)J\left(x,y\right)
\end{equation}
for any $x$ and $y$ in $\mathbb{R}$. We refer as $\bar{m}'$ to the derivative of $\bar{m}$ with respect to $x$.

\section{Uphill Diffusion}

\noindent
Our main result is
\begin{thm}\label{teo}
At fixed $\beta>1$ and $\mu\in\left(\ms,\mb\right)$, there is ${\ve}_{\beta}>0$ such that for any $\ve<{\ve}_{\beta}$ there are an antisymmetric, continuous function $m$ and a positive constant $j$ that solve
\begin{equation}\label{pbm}
\begin{dcases}
j\ve=-\chib\left(m\left(x\right)\right)\frac{\partial}{\partial x}\frac{\delta\mathcal{F}_{\beta}\left[m\mid \mu\right]
}{\delta m\left(x\right)},\qquad &x\in [-\vei,\vei] \\
m\hs(-\vei) =-\mu, \qquad m\hs(\vei) =\mu.
\end{dcases}
\end{equation}
\end{thm}

\subsection{Outline of the Proof}

\noindent
For our purposes, it is worth performing the following change of variables:
\begin{equation}
h\left(x\right)\coloneqq \frac{\delta\mathcal{F}_{\beta}\left[m\mid \mu\right]
}{\delta m\left(x\right)}
\end{equation}
where, explicitly
\begin{equation}\label{deltaf}
\frac{\delta\mathcal{F}_{\beta}\left[m\mid \mu\right]}{\delta m\left(x\right)}=
-\frac{1}{\beta}\,\mathrm{arc}\tanh\left(m\left(x\right)\right)+\int J\left(x,y\right)m\left(y\right) \de y.
\end{equation}
In this position \eqref{current} becomes, after a straight integration
\begin{equation}\label{chvar}
h\left(x\right)=h\left(x_0\right)-j\ve\int_{x_0}^x \frac{\de y}{\chib\left(m\left(y\right)\right)}.
\end{equation}
Observe that $h\left(0\right)=0$ if $m$ is odd, so we eventually formulate problem \eqref{pbm} as a system of coupled equations:
\begin{equation}\label{pbmh}
\begin{dcases}
m\left(x\right)= \tanh\Big\{\beta\Big[\left(J\ast m\right)\left(x\right)+h\left(x\right)
\Big]\Big\} \\
h\left(x\right)=-j\ve\int_0^x \frac{\de y}{\chib\left(m\left(y\right)\right)}
\end{dcases}
\qquad x\in [-\vei,\vei],
\end{equation}
with $m\hs(-\vei)=-\mu$, $m\hs(\vei)=\mu$. Notice that the first equation in \eqref{pbmh} is just \eqref{deltaf} expressed in the new variables. We will often indicate
\begin{equation}
\left[\mathsf{H}\left(m\right)\right]\left(x\right) \coloneqq -j\ve\int_0^x \frac{\de y}{\chib\left(m\left(y\right)\right)}\qquad \forall x\in[-\vei,\vei].
\end{equation}
The existence of a solution of problem \eqref{pbmh} is proved by iteration (Newton's method): we start from a couple $\left(m_0,h_0\right)$ and fixed $\mu$ and $j$ and define a feasible map $\left(m_{n},h_{n}\right)\mapsto\left(m_{n+1},h_{n+1}\right)$ that converges uniformly to a couple $\left(m,h\right)$ that solves \eqref{pbmh} and satisfies certain boundary conditions $m\hs(\pm\vei)=\pm\nu$, $\nu\neq\mu$ in general. Afterwards, we prove that $j$ can be actually tuned in order to cover the whole metastable region, that is for any $\nu\in\left(\ms,\mb\right)$ there exists at least one $j>0$ such that $\lim_{n\to\infty} m_n\left(x\right)$ solves \eqref{pbmh} with $m\,(\vei)=\nu$. In this scheme, the choice of $m_0$ (and $h_0$ as a function of $m_0$) turns out to be crucial, as we would like to start with a profile that is ``almost'' a fixed point.
\newline
\newline
The technical part of the paper is organized as follows: after having established the recursive method and chosen $m_0$, we perform in Section \ref{quattro} some estimates that are needed in the course of the proof; in particular, we prove the invertibility of $\mathsf{I}-\mathsf{A}_{m,h}$. In Section \ref{cinque} we construct the sequence $\left(m_{n},h_{n}\right)_{n=0}^{\infty}$ and prove convergence to a certain solution of \eqref{pbmh} with $j>0$. In Section \ref{sei} we deal with the invertibility issue mentioned above.

\subsection{Choice of $m_0$}

\begin{prop}
The ``macroscopic'' problem at $\beta>1$
\begin{equation}\label{pbmmacro}
\begin{dcases}
j_{M} = -\big[1-\chib\left(M\left(x\right)\right)\big]M'\left(x\right), \qquad x\in \left[0,1\right]\\
M\left(0\right)= \mu_-, \qquad M\left(1\right)=\mu_+
\end{dcases}
\end{equation}
with $0<\mu_+<\mu_-<1$, admits a unique solution in $C^{\infty}\left(\left[0,1\right]\right)$. Such solution is decreasing in $\left[0,1\right]$.
\end{prop}
\noindent
\textbf{Proof.} We refer to \eqref{pbmmacro} as the macroscopic equation because it comes from the variational problem that one obtains after performing the macroscopic limit (see \cite{RB2,RB}). A straight integration gives
\begin{equation}\label{terzogrado}
x=-\frac{1}{j_{M}}\Big[\left(\beta-1\right)\left(M\left(x\right)-\mu_-\right)-\frac{\beta^3}{3}\left(M^3\left(x\right)-\mu_-^3\right)\Big],
\end{equation}
with $j_M$ fixed by the choice of $\mu_-$ and $\mu_+$:
\begin{equation}
j_{M}=\left(\beta-1\right)\big(\mu_- -\mu_+\big)-\frac{\beta^3}{3}\big(\mu_-^3-\mu_+^3\big).
\end{equation}
As a function of $M$, $x$ is infinitely times differentiable and moreover, $x\left(M\right)$ is invertible since $M'$ is negative. $M$ can be obtained as the unique real solution of the cubic equation \eqref{terzogrado}.
\qed
\newline
\newline
Notice that problem \eqref{pbmmacro} can be formulated as a system of coupled equations as well
\begin{equation}\label{pbmhm}
\begin{dcases}
M\left(x\right)= \tanh\Big\{\beta\Big[M\left(x\right)+H\left(x\right)
\Big]\Big\} \\
H\left(x\right)=\widetilde{H}-j_{M}\int_0^x \frac{\de y}{\chib\left(M\left(y\right)\right)}
\end{dcases}
,\qquad x\in [0,1]
\end{equation}
with $\widetilde{H} =\beta^{-1}\mathrm{arc}\tanh\left(\mu_-\right)-\mu_-$. We define
\begin{equation}\label{m0}
m_0\left(x\right)\coloneqq
\begin{dcases}
\bar{m}\left(x\right) \qquad &0\le x\le\veim\\
M_{\mu}\left(\frac{\ve x-\sqve}{1-\sqve}\right)\eqqcolon \hat{M}_{\mu}\left(x\right) \qquad &\veim\le x\le\vei,
\end{dcases}
\end{equation}
with $m_0\left(x\right)=-m_0\left(-x\right)$ for $x<0,$ where $M_{\mu}$ is the solution of \eqref{pbmmacro} that satisfies $M_{\mu}\left(0\right)=\bar{m}\hs(\veim)$ and $M_{\mu}\left(1\right)=\mu$. For technical reasons, we choose $\vei$ so large that $\bar{m}\hs(\veim/2)=\mb-\delta$, $\delta>0$ a small parameter specified further on. We speculate that if $\vei$ is large enough, the solution should not differ so much from the instanton in the nearby of the origin. Once reached the value $m_0\hs(\veim)\approx\mb$, we suppose the solution to be monotone decreasing and ``close'' to the (rescaled) macroscopic profile. This will be very clear a posteriori, as we will show that in fact the distance between $m_0$ and the stationary solution $m$ is of order $\ve$ in the sup norm.

\subsection{Iterative scheme}

\noindent
The following results explicitly defines the method.
\begin{prop}\label{p3}
Let $m_0$ as in \eqref{m0} satisfying $m_0\,(\vei)=\mu_0$, $\mu_0\in\left(\ms,\mb\right)$ fixed and $h_0 = \mathsf{H}\left(m_0\right)$. For any $n$, there is $m_n\in C\left([-\vei,\vei],\left[-1,1\right]\right)$ that solves
\begin{equation}\label{mn}
\begin{dcases}
m_{n}\left(x\right)=\tanh\Big\{\beta\Big[\left(J\ast m_n\right)\left(x\right)+h_{n-1}\left(x\right)
\Big]\Big\}\\
h_{n-1}\left(x\right)=\left[\mathsf{H}\left(m_{n-1}\right)\right]\left(x\right)
\end{dcases}
\end{equation}
with $m_n\hs(-\vei)=-\mu_n$, $m_n\hs(\vei)=\mu_n$, $\mu_n\in\left(\ms,\mb\right)$, provided $\ve$ is small enough. The sequence $\left(m_n,h_n\right)_{n=0}^{\infty}$ converges uniformly to a pair $\left(m,h\right)$, where $h=\mathsf{H}\left(m\right)$, which is a solution of problem \eqref{pbmh} with boundary conditions $m\hs(-\vei)=-\mu$, $m\hs(\vei)=\mu$, $\mu$ in the metastable region. Then, $m$ also solves \eqref{pbm}.
\end{prop}

\begin{prop}\label{p4}
In the same hypothesis of Proposition \ref{p3}, for any $\mu\in\left(\ms,\mb\right)$ there is at least one $j>0$ such that the iterative scheme defined above converges to a solution of \eqref{pbm} with boundary conditions $m\hs(-\vei)=-\mu$, $m\hs(\vei)=\mu$.
\end{prop}

\noindent
In the iterations, $j=j\left(\mu_0\right)$ is fixed parameter, whose value is actually specified by the boundary value $m_0\,(\vei)=\mu_0$ (and $\bar{m}\,(\veim)$ that depends on $\ve$ only):
\begin{equation}
j=\left(\beta-1\right)\big(\bar{m}\hs(\veim) -\mu\big)-\frac{\beta^3}{3}\big(\bar{m}^3\hs(\veim)-\mu^3\big).
\end{equation}
Every time an iteration is performed, the boundary value changes, and therefore we cannot rule out the possibility that our constructive method defines a map $j\mapsto \left(\ms,\mb\right)$ which is not surjective. Hence, Proposition \ref{p4} is needed in order to close the proof of Theorem \ref{teo}.

\section{Some properties of $\mathsf{A}_{m,h}$}\label{quattro}

\subsection{A preliminary result}

\noindent
We recall here a result proved in \cite{DMOP} and that can be even found in \cite{EP}. Define the scalar product on $\mathbb{R}$
\begin{equation}
\left\langle f\right\rangle_{\infty}\coloneqq\int_{\mathbb{R}} f\left(x\right)\frac{\de x}{\bar{p}\left(x\right)}
\end{equation}
and indicate $\widetilde{m}'\coloneqq\bar{m}'/\sqrt{\left\langle (\bar{m}')^2\right\rangle_{\infty}}$. We have the following
\begin{prop}
There are positive constants $a$ and $c$ such that for any $f\in L^{\infty}\left(\mathbb{R}\right)$ and any integer $n$: 
\begin{equation}
\left|\,\int_{\mathbb{R}^n}{\mathsf{A}^n_{\bar{m}}}\left(x,y\right)\widetilde{f}\left(y\right)\de y\,\right| \le
c\hs\|\widetilde{f} \|_{\infty}\mathrm{e}^{-an}, \qquad \widetilde{f}\coloneqq f-\left\langle f\widetilde{{m}}'\right\rangle_{\infty}\widetilde{m}'.
\end{equation} 
\end{prop}
\noindent
There is a very straight consequence of this result, which is however essential for our purposes.
\begin{cor}
For any bounded, antisymmetric function $\psi$ on $\mathbb{R}$ and any integer $n$:
\begin{equation}\label{est}
\left|\,\int_{\mathbb{R}^n}{\mathsf{A}^n_{\bar{m}}}\left(x,y\right)\psi\left(y\right)\de y\,\right| \le c\hs\mathrm{e}^{-an} \left\|\psi\right\|_{\infty}.
\end{equation}
\end{cor}
\noindent
\textbf{Proof.} Since $\widetilde{m}'$ is symmetric, $\left\langle \psi\widetilde{{m}}'\right\rangle_{\infty}=0$ and then $\widetilde{\psi}\equiv\psi$.
\qed

\subsection{Invertibility of $\mathsf{I}-\mathsf{A}_{m,h}$}

\noindent
Define the set:
\begin{equation}
\Sigma_{\delta,\ve}\coloneqq\bigg\{m\in L^{\infty}\left([-\vei,\vei]\right) \mid \left\|p_{m,h}-p_{m_0,h_0}\right\|_{\ve}<\delta\bigg\}.
\end{equation}
\begin{lem}\label{lem11}
Let $m^{(1)},m^{(2)}\in\Sigma_{\delta,\ve}$. Then
\begin{equation}
\|\mathsf{H}(m^{(2)})-\mathsf{H}(m^{(1)})\|_{\ve}\le \tilde{c}\hs \|m^{(2)}-m^{(1)}\|_{\ve}, \qquad \tilde{c}= \frac{2\beta j}{\chib^2\left(\mb\right)}.
\end{equation}
\end{lem}
\noindent
\textbf{Proof.} We have, for any $x\in[-\vei,\vei]$:
\begin{eqnarray}
\big|[\mathsf{H}(m^{(2)})]\left(x\right)-[\mathsf{H}(m^{(1)})]\left(x\right)\big|&\le&
\beta j\ve \int_{0}^x \frac{\left|(m^{(2)})^2\left(y\right)-(m^{(1)})^2\left(y\right)\right|}{\chib(m^{(1)}\left(y\right))\chib(m^{(2)}\left(y\right))}\de y\nn\\
&\le& \frac{2\beta j}{\chib^2\left(\mb\right)}\|m^{(1)}-m^{(2)}\|_{\ve}.
\end{eqnarray}
\qed
\begin{comment}
\begin{lem}
Let $m\in\Sigma_{\delta,\ve}$, $h=\mathsf{H}\left(m\right)$. Then,
\begin{equation}\label{wil}
\left\|p_{m,h}-p_{m_0,h_0}\right\| \le 4\beta^2\left(1+\tilde{c}\right)\delta.
\end{equation}
\end{lem}
\noindent
\textbf{Proof.} For any $x_1$ and $x_2$ in $\mathbb{R}$
\begin{equation}\label{th}
\left|\tanh^2\left(x_2\right)-\tanh^2\left(x_1\right)\right|\le 4 \tanh\left|x_2-x_1\right|,
\end{equation}
therefore
\begin{equation}
\left|p_{m,h}\left(x\right)-p_{m_0,h_0}\left(x\right)\right|\le4\beta^2\left(\left|m\left(x\right)-m_0\left(x\right)\right|+\left|h\left(x\right)-h_0\left(x\right)\right|\right).
\end{equation}
Using Lemma \ref{lem11} we get \eqref{wil}.
\qed
\end{comment}
\begin{prop}\label{pe}
Let $m\in\Sigma_{\delta,\ve}$, $h=\mathsf{H}\left(m\right)$. There exists a constant $\theta=\theta\left(\delta,\ve\right)$ such that for any bounded, antisymmetric function $\psi$ on $\mathbb{R}$ and any integer $n$:
\begin{equation}\label{est1}
\left|\,\int_{|x_i|\le\veim,\, i=1,\ldots,n}\prod_{i=1}^n
\mathsf{A}_{m,h}\left(x_{i-1},x_i\right)\psi\left(x_n\right)\de x_1\ldots \de x_n\,\right| \le c\hs\mathrm{e}^{-an}\theta^n\sup_{|x_0|\le\veim}\left|\psi\left(x_0\right)\right|
\end{equation}
for any $\ve$ small enough.
\end{prop}
\noindent
\textbf{Proof.} Call, at fixed $x_0\in[-\veim,\veim]$:
\begin{equation}
\Delta_n^{(\psi)}\left(x_0\right)\coloneqq \int_{|x_i|\le\veim,\, i=1,\ldots,n}
\left|\prod_{i=1}^n \mathsf{A}_{m,h}\left(x_{i-1},x_i\right)-\prod_{i=1}^n {\mathsf{A}_{\bar{m}}}\left(x_{i-1},x_i\right)\right|\psi\left(x_n\right)\de x_1\ldots \de x_n.
\end{equation}
We have
\begin{eqnarray}
\Delta_n^{(\psi)}\left(x_0\right)&\le&\int_{|x_i|\le\veim,\, i=1,\ldots,n}\hspace*{-0.5mm}
\left|\prod_{i=1}^n \hspace*{-0.5mm}p_{m,h}\left(x_{i-1}\right)\hspace*{-0.75mm}J\left(x_{i-1},x_i\right)\hspace*{-0.5mm}-\hspace*{-0.5mm}
\prod_{i=1}^n \hspace*{-0.5mm} \bar{p}\left(x_{i-1}\right)\hspace*{-0.75mm}J\left(x_{i-1},x_i\right)\right|\hspace*{-0.75mm}\psi\left(x_n\right)\hspace*{-0.5mm}\de x_1\ldots \de x_n\nn\\
&\le& \int_{|x_i|\le\veim,\, i=1,\ldots,n}\hspace*{-0.5mm}
\left|\prod_{i=1}^n\hspace*{-0.5mm}\frac{p_{m,h}\left(x_{i-1}\right)}{\bar{p}\left(x_{i-1}\right)}-1\right|\prod_{i=1}^n {\mathsf{A}_{\bar{m}}}\left(x_{i-1},x_i\right)\psi\left(x_n\right)\de x_1\ldots\de x_n.
\end{eqnarray}
Write
\begin{equation}
\frac{p_{m,h}\left(x\right)}{\bar{p}\left(x\right)}=1+\frac{\left|p_{m,h}\left(x\right)-p_{m_0,h_0}\left(x\right)\right|}{\bar{p}\left(x\right)}+\frac{\left|p_{m_0,h_0}\left(x\right)-\bar{p}\left(x\right)\right|}{\bar{p}\left(x\right)}
\end{equation}
and observe that
\begin{equation}
\left|p_{m,h}\left(x\right)-p_{m_0,h_0}\left(x\right)\right|\le 4\beta^2\tilde{c}\sqve \qquad \mathrm{if}\, \left|x\right|\le\veim,
\end{equation}
where we used the fact that $\left|\tanh^2\left(x_2\right)-\tanh^2\left(x_1\right)\right|\le 4 \tanh\left|x_2-x_1\right|$ for any real numbers $x_1$ and $x_2$. Hence,
\begin{equation}
\sup_{\left|x\right|\le\veim}\frac{p_{m,h}\left(x\right)}{\bar{p}\left(x\right)}\le 1+\chib^{-1}\left(\mb\right)\left[\delta+4\beta^2\tilde{c}\sqve\right]\eqqcolon \theta.
\end{equation}
Therefore,
\begin{eqnarray}
\Delta_n^{(\psi)}\left(x_0\right)&\le&\left(\theta^n-1\right)\int_{|x_i|\le\veim,\, i=1,\ldots,n}
\prod_{i=1}^n {\mathsf{A}_{\bar{m}}}\left(x_{i-1},x_i\right)\psi\left(x_n\right)\de x_1\ldots\de x_n\nn\\
&\le&c\hs\mathrm{e}^{-an}\left(\theta^n-1\right)\sup_{|x_0|\le\veim}\left|\psi\left(x_0\right)\right|.
\end{eqnarray}
Combining the trivial inequality $|\mathsf{A}^n_{m,h}\psi|\le |{\mathsf{A}^n_{\bar{m}}}\psi|+|({\mathsf{A}^n_{m,h}}-{\mathsf{A}^n_{\bar{m}}})\psi|$ with \eqref{est}, we get the result. 
\qed
\begin{prop}\label{45}
Let $n_{\ve}\coloneqq \lfloor \veim -\frac{\veim}{2} \rfloor$ and $m\in\Sigma_{\delta,\ve}$, $h=\mathsf{H}\left(m\right)$. If $\ve$ is small enough, there is $0<\gamma<1$, $\gamma=\gamma\left(\delta,\ve\right)$, such that for any bounded, antisymmetric function $\psi$ on $[-\vei,\vei]$ and any $n\le n_{\ve}$:
\begin{equation}\label{crst}
\left|\,\int \prod_{i=1}^{n}
\mathsf{A}_{m,h}\left(x_{i-1},x_i\right)\psi\left(x_{n}\right)\de x_1\ldots \de x_{n}\,\right| \le 
\gamma^{n}\left\|\psi\right\|_{\ve}\qquad \forall x_0\in[-\vei,\vei].
\end{equation}
\end{prop}
\noindent
\textbf{Proof.} If $x_0\in[0,{\veim}/{2}]$, $\left|x_i\right|\le\veim$ for any $i=1,\ldots,n_{\ve}$ because $J$ has range $1$, thus estimate \eqref{est1} applies. If $x_0\in[{\veim}/{2},\vei]$, we distinguish two cases; if $\left|x_i\right|<\veim$, $\left|p_{m_0,h_0}\left(x_i\right)-\bar{p}\left(x_i\right)\right|\le 4\beta\tilde{c}\sqve$, while $\left|p_{m,h}\left(x_i\right)-p_{m_0,h_0}\left(x_i\right)\right|\le\delta$ by hypothesis. Since $\bar{p}\left(x_i\right)\le
\bar{p}\hs(\veim/2)=\chib\left(\mb\right)+\left(2\mb-\ve\right)\ve$, $\left|p_{m,h}\left(x_i\right)\right|\le\chib\left(\mb\right)+\delta+O\left(\ve\right)$, that is less than 1 for $\ve$ small enough. If $\left|x_i\right|>\veim$, we claim that $\left|p_{m_0,h_0}\left(x_i\right)-\chib\left(m_0\left(x_i\right)\right)\right|=O\left(\ve\right)$. Indeed,
\begin{eqnarray}\label{oggi}
\left|p_{m_0,h_0}\left(x_i\right)-\chib\left(m_0\left(x_i\right)\right)\right|&=\beta&\hs\left|\tanh^2\Big\{\beta\Big[\big(J\ast \hat{M}_{\mu}\big)\left(x\right)+h_0\left(x\right)\Big]\Big\} -\hat{M}^2_{\mu}\left(x\right)\right|\nn\\
&\le&4\beta^2 \left|\big(J\ast \hat{M}_{\mu}\big)\left(x\right)-\hat{M}_{\mu}\left(x\right)\right|\nn\\
&\le&4\beta^2 \max_{y:\hs\left|y-x\right|\le 1}\left|M_{\mu}\left(y\right)- M_{\mu}\left(x\right)\right|\nn\\
&\le&4\beta^2\left\|\frac{\de M_{\mu}}{\de x}\right\|_{\ve} = \mathrm{const}\,\cdot\,\ve,
\end{eqnarray}
then $\left|p_{m,h}\left(x_i\right)\right|\le\chib\left(\mu\right)+\delta+O\left(\ve\right)$. Therefore, if $\ve$ is small enough there exists a constant $\theta'<1$ that bounds $p_{m,h}\left(x_i\right)$ for any $\left|x_i\right|>\vei/2$. Define $\gamma$ as the maximum between $\theta$ and $\theta'$ to obtain \eqref{crst}.
\qed
\newline
\newline
Proposition \ref{45} induces the following
\begin{prop}\label{mj}
Let $m\in\Sigma_{\delta,\ve}$, $h=\mathsf{H}\left(m\right)$. For any bounded, antisymmetric function $\psi$ on $ [-\vei,\vei]$:
\begin{equation}\label{9-1}
\left|\sum_{n=0}^{\infty}\int\prod_{i=0}^n\mathsf{A}_{m,h}\left(x_{i-1},x_i\right)\psi\left(x_n\right)\de x_1\ldots \de x_n\right| \le \frac{\left\|\psi\right\|_{\ve}}{1-\gamma}
\end{equation}
if $\ve$ is small enough.
\end{prop}
\noindent
\textbf{Proof.} It suffices to show that for any integer $n$, $\big\|\mathsf{A}^n_{m,h}\psi\big\|_{\ve}\le\gamma^n \left\|\psi\right\|_{\ve}$. If $n\le n_{\ve}$ this is true because of \eqref{crst}. If $n>n_{\ve}$, we write $n=k_n n_{\ve} + m_n$, with $m< n_{\ve}$ so that
\begin{eqnarray}
\big\|\mathsf{A}^n_{m,h}\psi\big\|_{\ve}&\le&\sup_{\left|x_0\right|\le\vei}\bigg|\,
\int\prod_{i_1=1}^{n_{\ve}}\mathsf{A}_{m,h}\left(x_{i_1-1},x_{i_1}\right)\de x_1\ldots \de x_{n_{\ve}}\nn\\
&\times&\int\prod_{i_2=n_{\ve}+1}^{2n_{\ve}}\mathsf{A}_{m,h}\left(x_{i_2-1},x_{i_2}\right)\de x_{n_{\ve}+1}\ldots \de x_{2n_{\ve}}\nn\\
&\vdots&\nn\\
&\times&\int\prod_{i_k=\left(k-1\right)n_{\ve}+1}^{k n_{\ve}}\mathsf{A}_{m,h}\left(x_{i_k-1},x_{i_k}\right)
\de x_{\left(k-1\right)n_{\ve}+1}\ldots \de x_{k n_{\ve}}\nn\\
&\times&\int\prod_{j=k n_{\ve}+1}^{n}\mathsf{A}_{m,h}\left(x_{j-1},x_{j}\right)\psi\left(x_n\right)\de x_{k n_{\ve}+1}\ldots \de x_n\bigg|.
\end{eqnarray}
Call
\begin{equation}
\Psi^{(k)}\left(x_{k n_{\ve}}\right)\coloneqq\int\prod_{j=k n_{\ve}+1}^{n}\mathsf{A}_{m,h}\left(x_{j-1},x_{j}\right)\psi\left(x_n\right)\de x_{k n_{\ve}+1}\ldots \de x_n
\end{equation}
and notice that $\Psi^{(k)}$ is antisymmetric and satisfies $\|\Psi^{(k)}\|_{\ve}\le \gamma^{m_n}\left\|\psi\right\|_{\ve}$ by virtue of \eqref{crst}. Thus, for any $i<k$:
\begin{equation}
\Psi^{(i)}\left(x_{i n_{\ve}}\right)\coloneqq
\int\prod_{j=i n_{\ve}+1}^{\left(i+1\right)n_{\ve}}\mathsf{A}_{m,h}\left(x_{j-1},x_{j}\right)\Psi^{(i+1)}\left(x_n\right)\de x_{i n_{\ve}+1}\ldots \de x_{\left(i+1\right)n_{\ve}}
\end{equation}
is antisymmetric and therefore by iteration
\begin{eqnarray}
\big\|\mathsf{A}^n_{m_0,h_0}\psi\big\|_{\ve}&\le&\sup_{\left|x_0\right|\le\vei}\bigg|\,
\int\prod_{i_1=1}^{n_{\ve}}\mathsf{A}_{m,h}\left(x_{i_1-1},x_{i_1}\right)\Psi^{(1)}\left(x_{n_{\ve}}\right)\de x_1\ldots \de x_{n_{\ve}}\nn\\
&\le& \gamma^{n_{\ve}}\|\Psi^{(1)}\|_{\ve}\le \gamma^{2n_{\ve}}\|\Psi^{(2)}\|_{\ve}\le\ldots\le\gamma^{kn_{\ve}}\|\Psi^{(k)}\|_{\ve}\le \gamma^n\left\|\psi\right\|_{\ve}.
\end{eqnarray}
Summing on $k$ we get \eqref{9-1}. 
\qed
\newline
\newline
The previous bound induces the existence of the inverse of $\mathsf{I}-\mathsf{A}_{m,h}$. Explicitly
\begin{equation}
\left(\mathsf{I}-\mathsf{A}_{m,h}\right)^{-1}\coloneqq\sum_{k=0}^{\infty}\mathsf{A}^k_{m,h}, \qquad 
\|(\mathsf{I}-\mathsf{A}_{m,h})^{-1}\|_{\ve}\le \frac{1}{1-\gamma}.
\end{equation}
\begin{lem}\label{fund}
Let $m\in\Sigma_{\delta,\ve}$, $h=\mathsf{H}\left(m\right)$. For any bounded, antisymmetric function $F$ on $[-\vei,\vei]$ the equation
\begin{equation}
\vf\left(x\right)-p_{m,h}\left(x\right)\left(J\ast\vf\right)\left(x\right)=F\left(x\right)
\end{equation}
can be solved in the unknown function $\vf$. Furthermore, $\left\|\vf\right\|_{\ve}\le \left(1-\gamma\right)^{-1}\left\|F\right\|_{\ve}$.
\end{lem}
\noindent
\textbf{Proof.} It is a straightforward consequence of Proposition \ref{mj}.

\section{Newton's Method}\label{cinque}

\subsection{Small perturbations to $m_0$}

\noindent
In this section we construct $m_1$ as a series
\begin{equation}
m_1\left(x\right)=m_0\left(x\right)+\sum_{n=0}^{\infty}\vf_n\left(x\right)
\end{equation}
in which each correction $\vf_n$ depends on the previous ones
\begin{equation}
\vf_n\left(x\right)=\vf_n\left(x; m_0\left(x\right),\vf_1\left(x\right),\ldots,\vf_{n-1}\left(x\right)\right), \qquad n\ge1\\
\end{equation}
with $\vf_0\equiv 0$.
For notational convenience, we will often indicate
\begin{equation}
\phi_n\left(x\right)\coloneqq\sum_{m=1}^{n-1}\vf_m\left(x\right), \qquad n\ge 1.
\end{equation}

\begin{prop}
For any $\ve$ small enough, there exists an antisymmetric function $\vf_1\in L^{\infty}\left([-\vei,\vei]\right)$ that solves
\begin{equation}\label{vf1}
\begin{dcases}
\vf_1\left(x\right)-p_{m_0,h_0}\left(x\right)\left(J\ast\vf_1\right)\left(x\right)=\tanh\Big\{\beta\Big[\left(J\ast m_0\right)\left(x\right)+h_0\left(x\right)\Big]\Big\} -m_0\left(x\right)\\
\vf_1\hs(-\vei)=- t_1, \qquad \vf_1\hs(\vei)=t_1
\end{dcases}
\end{equation}
for some $t_1\in\mathbb{R}$. Moreover, there is $c_0>0$ such that $\left\|\vf_1\right\|_{\ve}\le c_0\ve$.
\end{prop}
\noindent
\textbf{Proof.} Call
\begin{equation}
S_0\left(x\right)\coloneqq \tanh\Big\{\beta\Big[\left(J\ast m_0\right)\left(x\right)+h_0\left(x\right)\Big]\Big\} -m_0\left(x\right).
\end{equation}
We restrict to the positive semiline. We have, for any $0\le x\le\veim-1$:
\begin{equation}
\left|S_0\left(x\right)\right|=\left|\tanh\Big\{\beta\Big[\left(J\ast \bar{m}\right)\left(x\right)+h_0\left(x\right)\Big]\Big\} -\bar{m}\left(x\right)\right|\le\beta j\ve \int_0^x \frac{\de y}{\chib\left(\bar{m}\left(y\right)\right)}\le\mathrm{const}\,\cdot\,\sqve.
\end{equation}
In the interval $\veim+1\le x \le\vei$, similarly to estimate \eqref{oggi}:
\begin{eqnarray}
\left|S_0\left(x\right)\right|&=&\left|\tanh\Big\{\beta\Big[\big(J\ast \hat{M}_{\mu}\big)\left(x\right)+h_0\left(x\right)\Big]\Big\} -\hat{M}_{\mu}\left(x\right)\right|\nn\\
&=&\left|\tanh\Big\{\beta\Big[\big(J\ast \hat{M}_{\mu}\big)\left(x\right)+h_0\left(x\right)\Big]\Big\} -
\tanh\Big\{\beta\Big[\hat{M}_{\mu}\left(x\right)+h_0\left(x\right)\Big]\Big\}
\right|\nn\\
&\le&\beta \left|\big(J\ast \hat{M}_{\mu}\big)\left(x\right)-\hat{M}_{\mu}\left(x\right)\right|\nn\\
&\le&\beta \max_{y:\hs\left|y-x\right|\le 1}\left|M_{\mu}\left(y\right)- M_{\mu}\left(x\right)\right|\nn\\
&\le&\beta\left\|\frac{\de M_{\mu}}{\de x}\right\|_{\ve} = \mathrm{const}\,\cdot\,\ve
\end{eqnarray}
where we used Lagrange's Theorem. The remaining case is when $\veim-1\le x\le\veim$. We consider at first the sub-case $\veim-1\le x\le\veim$. We get
\begin{eqnarray}
\left|S_0\left(x\right)\right|&=&\Bigg|
\tanh\Big\{\beta\Big[\int_{x-1}^{\veim}J\left(x,y\right)\bar{m}\left(y\right)\de y
+\int_{\veim}^{x+1}J\left(x,y\right)\hat{M}_{\mu}\left(y\right)\de y+h_0\left(x\right)\Big]\Big\}
-\bar{m}\left(x\right)\Bigg|\nn\\
&\le&\beta\left| \int_{\veim}^{x+1}J\left(x,y\right)\left(\hat{M}_{\mu}\left(y\right)-\bar{m}\left(y\right)\right)\de y +h_0\left(x\right)\right|\nn\\
&\le& \beta\left|\hat{M}_{\mu}\left(x+1\right)-\bar{m}\left(x+1\right)\right| + \mathrm{const}\,\cdot\,\sqve
\end{eqnarray}
because $\hat{M}_{\mu}$ is non increasing while $\bar{m}$ is increasing. Since by definition $\hat{M}_{\mu}\,(\veim)=\bar{m}\,(\veim)$:
\begin{equation}
\left|\hat{M}_{\mu}\left(x+1\right)-\bar{m}\left(x+1\right)\right|\le \left|\hat{M}_{\mu}\left(x+1\right)-\hat{M}_{\mu}\,(\veim)\right|+\left|\bar{m}\left(x+1\right)-\bar{m}\,(\veim)\right|=O\left(\ve\right)
\end{equation}
where we bounded the first term with the sup norm of the derivative of $\hat{M}_{\mu}$, while the second term is exponentially small in $\veim$. At last, if $\veim \le x\le\veim$:
\begin{eqnarray}
\left|S_0\left(x\right)\right|&=&\Bigg|
\tanh\Big\{\beta\Big[\int_{x-1}^{\veim}J\left(x,y\right)\bar{m}\left(y\right)\de y
+\int_{\veim}^{x+1}J\left(x,y\right)\hat{M}_{\mu}\left(y\right)\de y+h_0\left(x\right)\Big]\Big\}
-\hat{M}_{\mu}\left(x\right)\Bigg|\nn\\
&\le&\beta\left|
\int_{x-1}^{\veim}J\left(x,y\right)\left[\bar{m}\left(y\right)-\hat{M}_{\mu}\left(y\right)\right]\de y
+\int_{x-1}^{x+1}J\left(x,y\right)\hat{M}_{\mu}\left(y\right)\de y -\hat{M}_{\mu}\left(x\right)
\right|\nn\\
&\le&\beta\left|\bar{m}\left(x-1\right)-\hat{M}_{\mu}\left(x-1\right)\right|=O\left(\ve\right)
\end{eqnarray}
by virtue of the previous estimates. This proves that $\left\|S_0\right\|_{\ve}=O\left(\ve\right)$. Indeed, the existence of $\vf_1$ follows from the invertibility of $\mathsf{I}-\mathsf{A}_{m_0,h_0}$, and explicitly:
\begin{equation}
\vf_1\left(x\right)= \left(\mathsf{I}-\mathsf{A}_{m_0,h_0}\right)^{-1}S_0\left(x\right).
\end{equation}
By Lemma \ref{fund}, since $S\left(x\right)$ is antisymmetric, we get $\left\|\vf_1\right\|_{\ve}\le\left(1-\gamma\right)^{-1}\left\|S_0\right\|_{\ve}$.
\qed

\begin{prop}\label{5s}
For any $\ve$ small enough, we have
\begin{itemize}
\item[$\mathrm{(i)}$] for any $n\ge 2$ there exists an antisymmetric, bounded function $\vf_{n}\in L^{\infty}\left([-\vei,\vei]\right)$ that solves
\begin{equation}
\begin{dcases}
\vf_n\left(x\right)-p_{m_0+\phi_{n},h_0}\left(x\right)\left(J\ast\vf_n\right)\left(x\right)\hspace{-3mm}&=\tanh\Big\{\beta\Big[\left(J\ast m_0+\phi_n\right)\left(x\right)+h_0\left(x\right)\Big]\Big\}\nn\\
&-m_0\left(x\right)-\phi_n\left(x\right)\\
\vf_n\hs(-\vei)=-t_n, \qquad \vf_n\hs(\vei)=t_n
\end{dcases}
\end{equation}
for some $t_n\in\mathbb{R}$;
\item[$\mathrm{(ii)}$] there is a constant $\tau$ such that $\left\|\vf_n\right\|_{\ve}\le \tau \left\|\vf_{n-1}\right\|_{\ve}^2$ for any $n\ge 2$;
\item[$\mathrm{(iii)}$] $\lim_{n\to\infty}\left\|m_1-m_0-\phi_n\right\|_{\ve}=0$,
where $m_1$ is a solution of \eqref{mn} at $n=1$;
\item[$\mathrm{(iv)}$] $\left\|m_1-m_0\right\|_{\ve}=O\left(\ve\right)$.
\end{itemize}
\end{prop}
\noindent 
\textbf{Proof.} It works by induction. In particular, suppose that (i) and (ii) hold for any integer less or equal to a certain $k$. Since $\left\|\vf_1\right\|_{\ve}\le c_0\ve$, iterating (ii) we get
\begin{equation}
\left\|\vf_k\right\|_{\ve}\le \tau^{2^{k-1}-1}\left(c_0\ve\right)^{2^{k-1}}
\end{equation}
and then 
\begin{equation}\label{bd}
\left\|\phi_k\right\|_{\ve}\le 2c_0\ve
\end{equation}
provided $\ve\le\left(2c_0\tau\right)^{-2}$. Moreover if $\ve$ is so small that $m_0+\phi_k\in\Sigma_{\delta,\ve}$, and then $\mathsf{A}_{m_0+\phi_k,h_0}$ is invertible, and $\vf_{k+1}$ exists. We prove (ii); expand the hyperbolic tangent in Taylor series:
\begin{eqnarray}\label{expan}
\tanh\Big\{\beta\Big[\left(J\ast m_0+\phi_k\right)\left(x\right)+h_0\left(x\right)\Big]\Big\}&=&
\tanh\Big\{\beta\Big[\left(J\ast m_0+\phi_{k-1}\right)\left(x\right)+h_0\left(x\right)\Big]\Big\}\nn\\
&+&p_{m_0+\phi_{k-1},h_0}\left(x\right)\left(J\ast\vf_{k-1}\right)\left(x\right)\nn\\
&+&p'_{m_0+\phi_{k-1},h_0}\left(x\right)\left(J\ast\vf_{k-1}\right)^2\left(x\right)+\ldots .
\end{eqnarray}
Combining \eqref{expan} with the definition of $\vf_n$ we get
\begin{equation}
\vf_k\left(x\right)=\left(\mathsf{I}-\mathsf{A}_{m_0,\phi_{k-1},h_0}\right)^{-1}\left(
p'_{m_0,\phi_{k-1},h_0}\left(x\right)\left(J\ast\vf_{k-1}\right)\left(x\right)+\ldots\right),
\end{equation}
hence
\begin{equation}
\left\|\vf_k\right\|_{\ve}\le \left(1-\gamma\right)^{-1}\sup_{0\le x'\le \veim}p'_{m_0,\phi_{k-1},h_0}(x')\sup_{0\le x''\le \veim}\vf_{k-1}^2(x'')\le \frac{\beta}{1-\gamma}\left\|\vf_{k-1}\right\|^2_{\ve},
\end{equation}
so we can identify $\tau\equiv\beta/1-\gamma$. This proves that for any integer $n$:
\begin{equation}
m_0\left(x\right)+\phi_{n}\left(x\right)=\tanh\Big\{\beta\Big[\left(J\ast m_0+\phi_n\right)\left(x\right)+h_0\left(x\right)\Big]\Big\}+ O\left(\vf_n\left(x\right)\right).
\end{equation}
Taking the limit $n\to\infty$, by continuity of the hyperbolic tangent, we get the uniform convergence to $m_1$. (iv) follows from \eqref{bd}.
\qed

\subsection{Small perturbations to $h_0$}

\begin{prop}\label{55}
There is $\delta'>0$ such that for any $h\in C\left([-\vei,\vei]\right)$, if $\left\|h-h_0\right\|_{\ve}\le\delta'$:
\begin{itemize}
\item[$\mathrm{(i)}$] there exists a continuous, antisymmetric function $m$ solution of
\begin{equation}
m\left(x\right)= \tanh\Big\{\beta\Big[\left(J\ast m\right)\left(x\right)+h\left(x\right)\Big]\Big\};
\end{equation}
\item[$\mathrm{(ii)}$] $\left\|m-m_0\right\|_{\ve}=O\left(\delta'\right)$.
\end{itemize}
\end{prop}
\noindent
\textbf{Proof.} We construct $m$ using again Newton's method with starting point $\left(m_0,h\right)$. Observe that
\begin{equation}
\left\|p_{m_0,h}-p_{m_0,h_0}\right\|_{\ve}\le 4\beta\tanh\Big\{\beta \left\|h-h_0\right\|_{\ve}\Big\}\le 4\beta^2\delta';
\end{equation}
hence, if $\delta'<\delta/4\beta^2$ we are in the hypothesis of Lemma \ref{fund}, so we conclude that there exists an antisymmetric function $\psi_1$ that solves
\begin{equation}\label{vf1}
\begin{dcases}
\psi_1\left(x\right)-p_{m_0,h}\left(x\right)\left(J\ast\psi_1\right)\left(x\right)=\tanh\Big\{\beta\Big[\left(J\ast m_0\right)\left(x\right)+h\left(x\right)\Big]\Big\} -m_0\left(x\right)\\
\psi_1\hs(-\vei)=-s_1, \qquad \psi_1\hs(\vei)=s_1
\end{dcases}
\end{equation}
for some $s_1\in\mathbb{R}$, and moreover $\left\|\psi_1\right\|_{\ve}\le \tau\left\|h-h_0\right\|_{\ve}$. The rest of the proof is the same as that of Proposition \ref{5s}, provided $\delta'\le\delta/\left(2\tau+1\right)$, which is the condition needed in order to apply Lemma \ref{fund} recursively.
\qed

\subsection{Further corrections}

\noindent
As we shall see, it is worth emphasizing scaling properties of the magnetization profile by introducing the following weighted norm at fixed $\alpha>0$:
\begin{equation}\label{alpha}
\left\|m\right\|_{\ve,\alpha}\coloneqq \sup_{\left|x\right|\le \vei}\mathrm{e}^{-\alpha\ve \left|x\right|}\left|m\left(x\right)\right|, \qquad
m\in L^{\infty}\left([-\vei,\vei]\right).
\end{equation}
Notice that convergence in the $\alpha$-norm implies uniform convergence as the inclusion $\left\|\,\cdot\,\right\|_{\ve,\alpha}\le\left\|\,\cdot\,\right\|_{\ve}\le\mathrm{e}^{\alpha}\left\|\,\cdot\,\right\|_{\ve,\alpha}$ holds. The iterability of our method directly follows from the fact that $\left(m_n,h_n\right)\mapsto \left(m_{n+1},h_{n+1}\right)$ is a contraction in the $\alpha$-norm (for a feasible choice of parameters).

\begin{lem}\label{sera}
Let $h^{(i)}$, $i=1,2$ such that $\|h^{(i)}-h_0\|_{\ve}\le \delta'$, and $m^{(i)}$ the corresponding profiles constructed via Newton's method that solve
\begin{equation}
m^{(i)}\left(x\right)= \tanh\Big\{\beta\Big[\big(J\ast m^{(i)}\big)\left(x\right)+h^{(i)}\left(x\right)\Big]\Big\}.
\end{equation}
Then,
\begin{equation}
\|m^{(2)}-m^{(1)}\|_{\ve,\alpha}\le \tau \|h^{(2)}-h^{(1)}\|_{\ve,\alpha}
\end{equation}
for $\ve$ small enough.
\end{lem}
\noindent
\textbf{Proof.} $m^{(1)}$ and $m^{(2)}$ exist by Proposition \ref{55}. By Taylor's Theorem
\begin{equation}\label{6s}
m^{(2)}\left(x\right)= \tanh\Big\{\beta\Big[\big(J\ast m^{(1)}\big)\left(x\right)+h^{(1)}\left(x\right)\Big]\Big\}+R_{1,2}\left(x\right),
\end{equation}
where
\begin{equation}
R_{1,2}\left(x\right)=p_{1,2}\left(x\right)\left[\big(J\ast \big(m^{(2)}-m^{(1)}\big)\big)\left(x\right)+h^{(2)}\left(x\right)-h^{(1)}\left(x\right)\right]
\end{equation}
$p_{1,2}$ being some interpolating function between $p_{m^{(2)},h^{(2)}}$ and $p_{m^{(1)},h^{(1)}}$. We multiply by $\mathrm{e}^{-\alpha\ve \left|x\right|}$ equation \eqref{6s} and take absolute values to get
\begin{equation}\label{eqineq}
\mathrm{e}^{-\alpha\ve \left|x\right|}\big|m^{(2)}\left(x\right)-m^{(1)}\left(x\right)\big|-p_{1,2}\left(x\right)\mathrm{e}^{-\alpha\ve}\big(J\ast \big|m^{(2)}-m^{(1)}\big|\big)\left(x\right)\nn\\
\le p_{1,2}\left(x\right)\mathrm{e}^{-\alpha\ve \left|x\right|}\big|h^{(2)}\left(x\right)-h^{(1)}\left(x\right)\big|.
\end{equation}
\eqref{eqineq} as an equality can be solved in the unknown function $\mathrm{e}^{-\alpha\ve \left|x\right|}|m^{(2)}\left(x\right)-m^{(1)}\left(x\right)|$ provided $\|p_{1,2}\,\mathrm{e}^{\alpha\ve}-p_{m_0,h_0}\|_{\ve}\le\delta$. In this case, i.e. if $\ve$ is small enough:
\begin{equation}
\|m^{(2)}-m^{(1)}\|_{\ve,\alpha}\le\frac{\left\|p_{1,2}\right\|_{\ve}}{1-\gamma}\|h^{(2)}-h^{(1)}\|_{\ve,\alpha}\le \tau\|h^{(2)}-h^{(1)}\|_{\ve,\alpha}.
\end{equation}
\qed

\subsection{Convergence to $\left(m,h\right)$}

\noindent
We now prove Proposition \ref{p3}.
\newline
\newline
\textbf{Proof.} Suppose that for any $k<n$, $n$ a fixed integer, the following hypothesis hold true:
\begin{itemize}
\item[(H1)] there is a continuous, antisymmetric function $m_k$ which solves
\begin{equation}
m_k\left(x\right)= \tanh\Big\{\beta\Big[\left(J\ast m_k\right)\left(x\right)+h_{k-1}\left(x\right)\Big]\Big\}
\end{equation}
with boundary conditions $m_k\hs(-\vei)=-mu_k$, $m_k\hs(\vei)=mu_k$, $\mu_k\in(\ms,\mb)$, where $h_{k}=\mathsf{H}\left(m_k\right)$.
\item[(H2)] there is a constant $\rho\in\left(0,1\right)$ independent of $k$ such that
\begin{equation}
\left\|h_k-h_{k-1}\right\|_{\ve,\alpha}\le\rho \left\|h_{k-1}-h_{k-2}\right\|_{\ve,\alpha}, \qquad 2\le k<n.
\end{equation}
\end{itemize}
Notice that (H1) and (H2) imply:
\begin{eqnarray}
&&\left\|h_k-h_{k-1}\right\|_{\ve,\alpha}\le \frac{\rho}{1-\rho} \left\|h_1-h_{0}\right\|_{\ve,\alpha}\\
\nn\\
&&\left\|h_k-h_{0}\right\|_{\ve,\alpha}\le \frac{\rho}{1-\rho} \left\|h_1-h_{0}\right\|_{\ve,\alpha}. \label{h4}
\end{eqnarray}
According to \eqref{h4}, $\alpha$ and $\ve$ can be suitably tuned in order to apply Lemma \ref{sera}; in particular, this is true uniformly in $k$ provided $\left\|h_1-h_0\right\|_{\ve}\le \left(1-\rho\right)/\mathrm{e}^{\alpha}\rho$. In this hypothesis there exists $m_n$ solution of 
\begin{equation}
m_n\left(x\right)= \tanh\Big\{\beta\Big[\left(J\ast m_n\right)\left(x\right)+h_{n-1}\left(x\right)\Big]\Big\}
\end{equation}
satisfying a certain boundary condition. It remains to prove that (H2) holds for $k=n$. We have, for any $x\in[-\vei,\vei]$:
\begin{eqnarray}
\left|h_n\left(x\right)-h_{n-1}\left(x\right)\right|&\le&2\beta\tilde{c}\,\ve
\int_0^x \mathrm{e}^{-\alpha\ve y} \left|m_n\left(x\right)-m_{n-1}\left(x\right)\right|\mathrm{e}^{\alpha\ve y}\,\de y\nn\\
&\le&2\beta\tilde{c}\,\ve\left\|m_n-m_{n-1}\right\|_{\ve,\alpha}
\int_0^x \mathrm{e}^{\alpha\ve y} \de y\nn\\
&\le&\frac{2\beta\tilde{c}}{\alpha}\left\|m_n-m_{n-1}\right\|_{\ve,\alpha}.
\end{eqnarray}
Hence, in the $\alpha$-norm:
\begin{equation}
\left\|h_n-h_{n-1}\right\|_{\ve,\alpha}\le \frac{2\beta\tilde{c}}{\alpha}\left\|m_n-m_{n-1}\right\|_{\ve,\alpha}\le\frac{2\beta\tilde{c}\,\tau }{\alpha}\left\|h_{n-1}-h_{n-2}\right\|_{\ve,\alpha}.
\end{equation}
If $\rho\le 2\beta\tilde{c}\,\tau/\alpha$ and $\ve$ is accordingly small, $\left(m_n,h_n\right)\mapsto\left(m_{n+1},h_{n+1}\right)$ is a contraction in the $\alpha$-norm. This implies uniform convergence to a solution of \eqref{pbmh}.
\qed

\section{Invertibility of the method}\label{sei}

\noindent
We indicate in this section $m_0\left(\,\cdot\,,j\right)$ as the starting magnetization profile carrying current $j$. This also fix the boundary value $m_0\hs(\vei)=\mu_0$. We use the same notation for the corresponding auxiliary magnetic field.

\subsection{Lipschitz continuity in $j$}

\begin{prop}
$m_0$ and $h_0$ are Lipschitz continuous in the current $j$.
\end{prop}
\noindent
\textbf{Proof.} We differentiate $m_0$ and $h_0$ with respect to $j$:
\begin{eqnarray}
\frac{\partial m_0}{\partial j}\left(x,j\right)&=&
\begin{dcases}
0 &0\le x\le\veim \\
-\frac{1}{1-\chib\left(m_0\left(x,j\right)\right)}\frac{\ve x-\sqve}{1-\sqve} &\veim\le x\le\vei,
\end{dcases}
\\ \nn \\ \nn \\
\frac{\partial h_0}{\partial j}\left(x,j\right)&=&
\begin{dcases}
0 &0\le x \le\veim \\
-\frac{1}{\chib\left(m_0\left(x,j\right)\right)}\frac{\ve x-\sqve}{1-\sqve} &\veim \le x\le\vei
\end{dcases}
\end{eqnarray}
and similarly for $x<0$. Then, for any $j_1$ and $j_2$, using Lagrange's Theorem:
\begin{eqnarray}
&&\left\|m_0\left(\,\cdot\,,j_2\right)-m_0\left(\,\cdot\,,j_1\right)\right\|_{\ve}\le \left(1-\chib\left(\ms\right)\right)^{-1}\left|j_2-j_1\right|\\ 
\nn \\
&&\left\|h_0\left(\,\cdot\,,j_2\right)-h_0\left(\,\cdot\,,j_1\right)\right\|_{\ve}\le \chib^{-1}\left(\mb\right)\left|j_2-j_1\right|.
\end{eqnarray}
\qed
\begin{prop}
The sequence $\left(m_n\left(\,\cdot\,,j\right)\right)_{n=0}^{\infty}$ is Lipschitz continuous in $j$. The limit profile $m\left(\,\cdot\,,j\right)$ is Lipschitz in $j$ as well.
\end{prop}
\noindent
\textbf{Proof.} Suppose that
\begin{itemize}
\item[(H)] There is $\rho'\in\left(0,1\right)$ such that, for any $k<n$, $n$ fixed:
\begin{equation}
\left\|h_k\left(\,\cdot\,,j_2\right)-h_k\left(\,\cdot\,,j_1\right)\right\|_{\ve,\alpha}\le \rho'
\left\|h_{k-1}\left(\,\cdot\,,j_2\right)-h_{k-1}\left(\,\cdot\,,j_1\right)\right\|_{\ve,\alpha}.
\end{equation}
\end{itemize}
It can be proved (in a way similar at all to that used to prove Lemma \ref{sera}) that (H) implies
\begin{equation}\label{7set}
\left\|m_{k+1}\left(\,\cdot\,,j_2\right)-m_{k+1}\left(\,\cdot\,,j_1\right)\right\|_{\ve,\alpha}\le \rho'
\left\|h_k\left(\,\cdot\,,j_2\right)-h_k\left(\,\cdot\,,j_1\right)\right\|_{\ve,\alpha}
\end{equation}
and therefore, there exists a constant $L>0$ such that for any $x\in[-\vei,\vei]$:
\begin{equation}
\left|h_{k+1}\left(x,j_2\right)-h_{k+1}\left(x,j_1\right)\right|\le\frac{L\mathrm{e}^{\alpha\ve \left|x\right|}}{\alpha}
\left\|m_{k+1}\left(\,\cdot\,,j_2\right)-m_{k+1}\left(\,\cdot\,,j_1\right)\right\|_{\ve,\alpha}.
\end{equation}
Multiplying both members by $\mathrm{e}^{-\alpha\ve \left|x\right|}$ and taking the supremum with respect to $x$ we get the inequality in the $\alpha$-norm:
\begin{eqnarray}\label{rec}
\left\|h_{k+1}\left(\,\cdot\,,j_2\right)-h_{k+1}\left(\,\cdot\,,j_1\right)\right\|_{\ve,\alpha}&\le& \frac{L}{\alpha}\left\|m_{k+1}\left(\,\cdot\,,j_2\right)-m_{k+1}\left(\,\cdot\,,j_1\right)\right\|_{\ve,\alpha}\nn\\
&\le&\frac{L\rho'}{\alpha}\left\|h_{k+1}\left(\,\cdot\,,j_2\right)-h_{k+1}\left(\,\cdot\,,j_1\right)\right\|_{\ve,\alpha}.
\end{eqnarray}
Choose $\alpha>L\rho'$ and let $\rho'\equiv L\rho'/\alpha<1$; then, (H) also holds for $n+1$. Moreover we get, applying recursively \eqref{rec}:
\begin{eqnarray}
\left\|h_{k}\left(\,\cdot\,,j_2\right)-h_{k}\left(\,\cdot\,,j_1\right)\right\|_{\ve,\alpha}&\le&\frac{\rho'}{1-\rho'}\left\|h_{0}\left(\,\cdot\,,j_2\right)-h_{0}\left(\,\cdot\,,j_1\right)\right\|_{\ve,\alpha}\nn\\
&\le& \frac{\rho'}{\left(1-\rho'\right)\chib\left(\mb\right)}\left|j_2-j_1\right|
\end{eqnarray}
and then, by \eqref{7set}:
\begin{equation}
\left\|m_{k}\left(\,\cdot\,,j_2\right)-m_{k}\left(\,\cdot\,,j_1\right)\right\|_{\ve,\alpha}\le
\frac{\rho'}{\left(1-\rho'\right)\chib\left(\mb\right)}\left|j_2-j_1\right|.
\end{equation}
The fact that the function $m$ that is obtained as limit of Newton's method is Lipschitz is a consequence of the uniform continuity of the sequence.
\qed

\subsection{Proof of Theorem \ref{teo}}

\noindent
What remains to prove is the surjectivity of $j$ as a function of $\mu_0$.
\begin{prop}
At fixed $\mu_0\in\left(\ms,\mb\right)$, let $\mu =m\,(\vei,j\left(\mu_0\right))$ the right boundary condition satisfied by the limit profile of the sequence starting with $m_0\left(\,\cdot\,,j\left(\mu_0\right)\right)$. For any $\mu_0\in\left(\ms,\mb\right)$ there are $j^-\left(\mu_0\right)$ and $j^+\left(\mu_0\right)$ such that
\begin{equation}
m\,(\vei,j^-\left(\mu_0\right))=\mu^-, \qquad m\,(\vei,j^+\left(\mu_0\right))=\mu^+
\end{equation}
with $\mu^-<\mu<\mu^+$.
\end{prop}
\noindent
\textbf{Proof.} This result closes the proof of theorem \ref{teo} as it provides sufficient conditions in order to apply the intermediate value theorem. Indeed, set
\begin{equation}
j^{\pm}_{\eta}\left(\mu_0\right)\coloneqq j\left(\mu_0\right)\mp\left[\left(\chib\left(\mu_0\right)-1\right)\eta + \beta\mu_0\eta^2 -\frac{\beta}{3}\eta^3\right],
\end{equation}
which are the currents corresponding to the starting magnetization value $\mu_0\pm\eta$ (notice that $j^{-}_{\eta}\left(\mu_0\right)>j^{+}_{\eta}\left(\mu_0\right)$). We have proved that $\left|\mu-\mu_0\right|=O\left(\ve\right)$, thus we can choose $\ve$ so small (depending on $\eta$ fixed) such that $\mu^-$ is smaller than $\mu$ while $\mu^+$ is larger than $\mu$. Since $m\left(\,\cdot\,,j\right)$ is continuous in $j$, the intermediate value theorem can be applied. Then, $m\,(\vei,j)$ takes any value between $m\,(\vei,j_{\eta}^+\left(\mu_0\right))$ and $m\,(\vei,j_{\eta}^-\left(\mu_0\right))$.
\qed

\section*{Acknowledgements}

\noindent
We thank Anna De Masi, Errico Presutti and Dimitrios Tsagkarogiannis for enlightening discussions.

%%%%%%%%%%%%%%%%%%%%%%%%%%%%%%%%%%%%%%%%%%%%%%%%%%%%%%

\end{document}